\documentclass[12pt, a4paper]{article}

\usepackage[utf8]{inputenc}    
\usepackage{graphicx}         
\usepackage{amsmath}           
\usepackage{amssymb}           
\usepackage{setspace}          
\usepackage{hyperref}          
\usepackage{authblk}           
\usepackage{longtable}         
\usepackage{booktabs}          
\usepackage{caption}           
\usepackage{multirow}          
\usepackage{xcolor}           
\usepackage{float}            
\usepackage{tabularx}
\usepackage{threeparttable}
\usepackage{booktabs}
\usepackage{float}
\usepackage{array}  
\newcolumntype{C}[1]{>{\centering\arraybackslash}p{#1}} 
\newcolumntype{P}[1]{>{\centering\arraybackslash}p{#1}}
\newcolumntype{M}[1]{>{\centering\arraybackslash}m{#1}}

\usepackage[a4paper, margin=1in]{geometry}

\usepackage[backend=biber, style=apa]{biblatex}
\usepackage{newtxtext}
\usepackage{newtxmath}

\begin{document}	
\title{Not Just Large: Tall Teams Dominate East Asia's Scientific Production}
\author{
    Siyuan Liu\textsuperscript{1},
    Wenjin Xie\textsuperscript{1},
    Wenyu Chen\textsuperscript{2},
    Tao Jia\textsuperscript{1,3}\thanks{Corresponding author: Tao Jia, (Email: tjia@swu.edu.cn; ORCID: 0000-0002-2337-2857).}\\
    \textsuperscript{1}College of Computer and Information Science, Southwest University, Chongqing, 400715, China \\
    \textsuperscript{2}College of Science, Guangxi University of Science and Technology, Liuzhou, 545006, China \\
    \textsuperscript{3}College of Computer and Information Science, Chongqing Normal University, Chongqing, 400047, China 
}

\maketitle
	
\begin{abstract}

\textbf{Purpose}: This study compares the hierarchical structure of scientific teams across countries and investigates factors associated with the observed cross-national differences.

\textbf{Design\slash methodology\slash approach}\textbf{:} Drawing on 150,817 publications with author contribution statements, we focus on the 15 countries with the largest volume of scientific publications, examine cross-country variations in the proportion of tall teams, and analyze how this proportion correlates with other factors.

\textbf{Findings}: Scientific output from East Asia is dominated by tall teams, which persist after controlling for team size, indicating that this pattern cannot be fully accounted for by the prevalence of larger teams in these countries. Cultural factors, measured by Power Distance, as well as the observed funding patterns of major basic science agencies, are associated with the dominance of tall teams in East Asia.

\textbf{Research limitations}: This study is limited by its reliance on publications with author contribution statements, which may introduce selection bias; its focus on cultural and funding factors, while leaving other institutional contexts unexamined; and its use of a leadership concentration measure that does not capture other dimensions of hierarchy.

\textbf{Practical implications}: Understanding cross-national differences in research team structures and their associated cultural and institutional factors can inform science policy and team management.

\textbf{Originality/value}: This study provides a systematic cross-national comparison of team hierarchy and offers a mechanistic understanding of the dominance of tall teams in East Asia, highlighting associations with cultural and funding factors.

\end{abstract}
	
\noindent \textbf{Keywords}: Scientific leadership; Team size; Team hierarchy; Social culture; Research funding

\section{Introduction}

Scientific collaboration is the primary driving force behind innovation and development in modern science \parencite{leahey2016sole, zeng2021fresh, wu2019large, wang2020science, lin2023remote, li2023author}. With the growing specialization of knowledge \parencite{jones2009burden} and the need to integrate diverse expertises to solve complex research problems \parencite{katz1997research}, scientific research has become increasingly team-oriented, characterized by both a rising share of output generated by teams and a steady growth in team size \parencite{wuchty2007increasing,fortunato2018science,lariviere2015team,li2025determination, hu2020mapping}.

The scientific team becomes not only larger, but also increasingly organized \parencite{walsh2015bureaucratization}. Leading a scientific team is like running a small business, in which each member holds a unique role according to their specialization and administrative level \parencite{carayol2004does,haeussler2020division,hays2015not,anderson2010functions,halevy2011functional}. In such organized settings, hierarchical structure is critical because it fundamentally shapes team processes and, consequently, performance \parencite{greer2018and}. Classical theories propose competing mechanisms for this influence: it can enhance coordination and efficiency \parencite{halevy2011functional,keltner2008reciprocal,van2009hierarchy}, or it may stifle communication and heighten conflict \parencite{bloom1999performance,greer2010equality,greer2017dysfunctions,wolfe2005perceived}. Among its various dimensions, the concentration of leadership, ranging from the tall team with centralized leadership to the flat team with shared responsibilities, has been shown to be associated with scientific outcomes. Specifically, the tall team tends to build on existing ideas and achieve short-term impact, whereas the flat team is more likely to generate disruptive, long-term innovations \parencite{xu2022flat}. Therefore, maintaining a balance between the tall team and the flat team is critical for sustaining both incremental and breakthrough advances.

A recent study shows that scientific output in East Asia, particularly in China, Japan, and South Korea, is dominated by large teams \parencite{liu2021dominance}.  Since team size and team hierarchy are often correlated \parencite{xu2022flat}, it remains unclear whether East Asian teams are also characterized by pronounced hierarchy. To investigate this, we ask the following research questions:

Q1: Is scientific output in East Asia also dominated by tall teams?

Q2: If so, what factors are associated with this dominance? Specifically, is it primarily explained by the prevalence of large teams, or is it associated with other factors such as social culture and funding patterns?

To answer these questions, we conduct a systematic empirical analysis of team hierarchy across the 15 countries with the highest publication output, using author contribution statements from 150,817 papers published in PLOS and PNAS between 2006 and 2019. China, Japan, and South Korea, the only East Asian countries among the 15 analyzed, share well-documented cultural foundations of hierarchical relationalism and authority systems rooted in Confucian and Buddhist traditions, providing a theoretically meaningful basis for analyzing them as a coherent group when studying team hierarchy \parencite{liu2015globalizing, hamilton1988market}. Our analyses show that East Asian scientific output is disproportionately produced by tall teams, a pattern that cannot be fully explained by differences in team size alone. We therefore propose two factors that are theoretically and empirically closely related to team hierarchy. The first is the cultural context, motivated by the long-standing traditions of hierarchical relationalism in China, Japan, and South Korea, rooted in Confucian thought \parencite{liu2015globalizing,hamilton1988market}. The second factor is the funding pattern, motivated by prior work showing that in China, major basic-science funding agencies are associated with larger research teams \parencite{liu2021dominance}, which tend to have more hierarchical structures \parencite{xu2022flat}. We find that each factor individually accounts for a substantial portion of the hierarchy gap between East Asian and non-East Asian teams.

Our study makes two main contributions. First, we provide a systematic cross-national analysis of team hierarchy, revealing a distinctive East Asian pattern in both static and temporal dimensions. Second, we conduct a mechanism-oriented analysis, linking this distinctive hierarchy to national cultural factors and funding patterns, offering a mechanistic understanding of cross-national differences in team structure. Together, these findings provide data-driven insights for team formation and management across national contexts.

\section{Data and methods}

\subsection{Data set}\label{subsec2_1}

We focus on papers published in seven PLOS journals (PLOS ONE, PLOS Biology, PLOS Computational Biology, PLOS Genetics, PLOS Medicine, PLOS Pathogens, and PLOS Neglected Tropical Diseases) and in the Proceedings of the National Academy of Sciences (PNAS), which provide standardized author contribution statements \parencite{maddi2024beyond}. To obtain these papers, we use publication data from the OpenAlex dataset \parencite{priem2022openalex}, identifying relevant papers and their Digital Object Identifiers (DOIs). We then collect the corresponding author contribution statements from journal websites. In total, we obtain 150,817 papers with contribution statements, including 132,329 PLOS papers and 18,488 PNAS papers. We parse contribution statements to identify the roles of each author within a given research team using an automated text-parsing framework. Detailed parsing procedures are provided in Supplementary Section 1.1.

\subsection{Country allocation}

There are generally two ways to determine the country to which a paper belongs. Straight counting \parencite{huang2011counting,waltman2015field,zheng2014influences,gonzalez2017dominance,kahn2016return,mazloumian2013global} assigns a paper to the country of its first or corresponding affiliation, whereas whole counting or fractional counting \parencite{sivertsen2019measuring,larsen2008state,lin2013influences} assigns the paper to all participating countries. Since the latter approach can result in multiple counting issues, and a previous study \parencite{huang2011counting} suggests that straight counting may be more effective for assessing scientific output at the country level. We therefore adopt straight counting in this work. In addition, it is reported that a paper’s first affiliation and corresponding affiliation usually point to the same country in Web of Science (WoS) \parencite{liu2021dominance}. Consequently, we use straight counting based on the first affiliation to determine the country with which a paper is affiliated. To assess the potential bias arising from international collaborations, the analysis is restricted to papers authored solely by researchers from a single country. The full results of this robustness check are presented in Supplementary Section 4, which demonstrates that the main conclusions are unaffected.

\subsection{Countries considered}

We focus on the 15 countries with the highest paper volume. These countries are the United States (US), China (CN), Germany (DE), United Kingdom (GB), France (FR), Japan (JP), Canada (CA), Australia (AU), Netherlands (NL), Spain (ES), Italy (IT), Brazil (BR), Sweden (SE), South Korea (KR), and Switzerland (CH). These countries largely correspond to the top 15 countries in terms of total publications based on WoS records \parencite{liu2021dominance}. We consider papers from mainland China, Hong Kong, and Macau as China's scientific output. 

Although the term ``East Asia'' encompasses a broader set of nations, we find that China, Japan, and South Korea, together account for approximately 91\% of all publications from East Asia in our dataset. Therefore, we believe that focusing on these three countries can sufficiently represent East Asia's scientific output.

\subsection{Research field}

OpenAlex employs a paper-level, bottom-up classification of research fields by quantifying semantic similarities between textual content across publications. These similarities are organized into hierarchical concepts spanning six levels, with the highest level manually curated to ensure conceptual coherence. This framework classifies research into 19 fields \parencite{priem2022openalex}: Biology, Chemistry, Physics, Mathematics, Computer science, Engineering, Sociology, Psychology, Economics, Political science, History, Philosophy, Medicine, Geography, Business, Materials science, Environmental science, Geology, and Art. The corresponding publication counts for each country and field are provided in Supplementary Table S1.

\subsection{Tall teams and flat teams}

Team hierarchy is a multidimensional construct, primarily conceptualized through two lenses. The inequality lens focuses on differences in valued attributes like power or status, which includes two dimensions: centralization (the concentration of such attributes in one or a few members) \parencite{ahuja1999network,argote1989centralize,bunderson2003recognizing} and steepness (the magnitude of absolute differences among all members) \parencite{anderson2010functions}. The relational lens defines hierarchy as acyclicity in dyadic influence relations \parencite{bunderson2016different}. We focus on the concentration of leadership within teams, given its established association with scientific outcomes \parencite{xu2022flat}. Identifying lead authors is essential for measuring leadership concentration. While author order is often used, it may not reflect actual contributions due to disciplinary differences in authorship conventions. Recently, author contribution statements have become a more reliable means of identifying lead authors \parencite{lu2020co, lariviere2016contributorship, haeussler2020division}. Based on author contribution statements, we adopt a shared leadership definition \parencite{xu2022flat}, where lead authors are those contributing to at least one of the following roles (past tense in statements): ``conceived'', ``designed'', ``led'', ``supervised'', ``coordinated'', ``interpreted'', and ``wrote''. Using this definition, we operationalize leadership concentration via the $L$-ratio, defined as the proportion of lead authors among all team members \parencite{xu2022flat, zhao2024more}. For a paper with \( n \) authors, the \( L \)-ratio ranges from $\frac{1}{n}$ (one author leads the entire team, representing maximum hierarchy) to 1 (all authors equally leading, representing a fully flat team).

 We are aware about a stricter definition, thought leadership, in which lead authors are identified based on the contribution role explicitly stated as ``conceived and designed the experiments'' \parencite{zhao2024more}. We perform the analyses based on this quantification and present the results in Supplementary Section 3.  In general, our findings are not affected by the choice of team hierarchy measure. The detailed procedures for identifying lead authors under both definitions are provided in Supplementary Section 1.2.

In organizational research, tall teams are organizations with concentrated managerial authority, whereas flat teams are those with broader spans of control \parencite{carzo1969effects,ghiselli1972leadership,porter1964effects}. For analytical purposes, we operationalize this distinction using $L$-ratio thresholds. Although there is no widely established threshold for a tall team, we follow an existing work \parencite{xu2022flat} and classify a team with $L$-ratio $<$ 0.5 as tall. We also consider two alternative thresholds ($L$-ratio $<$ 0.4 and $L$-ratio $<$ 0.6) and find that our findings are not affected by the choice of the threshold value (Supplementary Section 2).

\subsection{Funding information}

We obtain funding information for individual papers by matching OpenAlex records with WoS via DOIs. WoS has recorded grant numbers and funding agencies since 2008 \parencite{alvarez2017funding, tang2017funding}. Due to inconsistent naming of funding agencies across publications, we manually verify the frequently occurring name variants of major funding agencies for each country. Specifically, we verify the major funding agencies in China (NSFC, MOST), Japan (JSPS, JST), South Korea (NRF, MEST), the United States (NSF, NIH), Germany (DFG), the United Kingdom (MRC, BBSRC, EPSRC, NERC, ESRC), France (ANR), Canada (NSERC), Switzerland (SNSF), Sweden (VR), the Netherlands (NWO), Australia (ARC), and Brazil (CNPq). The full official names of these agencies, the detailed verification procedure, and the complete, reproducible mapping table of funding-agency name variants are provided in Supplementary Section 1.3.

Within each country, many papers supported by a primary basic science funding agency also receive joint funding from other major agencies. For example, 24.33\% of NSFC-funded papers additionally receive support from MOST, which is associated with flatter teams among larger teams, and 22.56\% of NSF-funded papers also receive NIH support, favoring tall teams regardless of size (Supplementary Figure S1). To reduce potential biases from overlapping funding, we exclude papers that acknowledge both agencies within each country (NSFC and MOST in China, JSPS and JST in Japan, NRF and MEST in South Korea, NSF and NIH in the United States), retaining only papers primarily funded by NSFC, JSPS, NRF, or NSF for our analysis. The overlap ratios between paired agencies are listed in Supplementary Table S4.

\subsection{Power Distance Index}

The Power Distance Index (PDI) scores for the 15 countries analyzed in this study are based on the 2015 version of Hofstede's cultural dimension dataset, publicly accessible at \url{https://geerthofstede.com/research-and-vsm/dimension-data-matrix/}. Our study period spans 2009–2019, closely aligning with this dataset. Since a country’s societal culture is generally stable over short time spans and changes only gradually across generations, the 2015 scores provide a valid reference for assessing cultural influences during the study period.

\subsection{Regression framework}

To estimate the associations between team hierarchy and key explanatory factors while accounting for potential confounders, we employ paper-level ordinary least squares (OLS) regressions. We examine three explanatory dimensions separately: regional affiliation, societal hierarchy culture, and national funding patterns. Estimating each in its own regression model allows us to isolate their respective associations with hierarchy while avoiding multicollinearity concerns that would arise from including them jointly. For each dimension, we estimate the following model:
\begin{equation}
L_i = \alpha + \beta_1 X_i + \beta_2 m_i + \beta_3 (X_i \times m_i) + \gamma C_i + \delta_\text{Year} + \delta_\text{Field} + \epsilon_i,
\end{equation}
where $L_i$ is the $L$-ratio of paper $i$, $X_i$ denotes the dimension-specific explanatory variable: for the regional dimension, $X_i$ is the EastAsia dummy (EA), which equals 1 if paper $i$ is produced by a team from China, Japan, or South Korea and 0 otherwise; for the cultural dimension, $X_i$ is the country-level PDI; and for the funding dimension, $X_i$ measures the proportion of tall-team papers funded by the country's primary national basic science funding agency ($f$). $m_i$ is team size, and the interaction term $X_i \times m_i$ tests whether the relationship between team size and hierarchy is moderated by the explanatory variable $X_i$. The vector $C_i$ includes paper-level controls: citation impact ($c_{5}$, citations within five years) and InternationalCollaboration (a binary variable indicating whether the paper involves authors from multiple countries). All models include year and field fixed effects ($\delta_\text{Year}$, $\delta_\text{Field}$), with standard errors clustered at the country level. Continuous explanatory variables ($m_i$, PDI, and $f$) are standardized using z-scores prior to estimation.

\section{Results}

\subsection{Tall-team dominance in East Asian countries}

\begin{figure}[h]
  \centering
  \includegraphics[width = 0.86\textwidth]{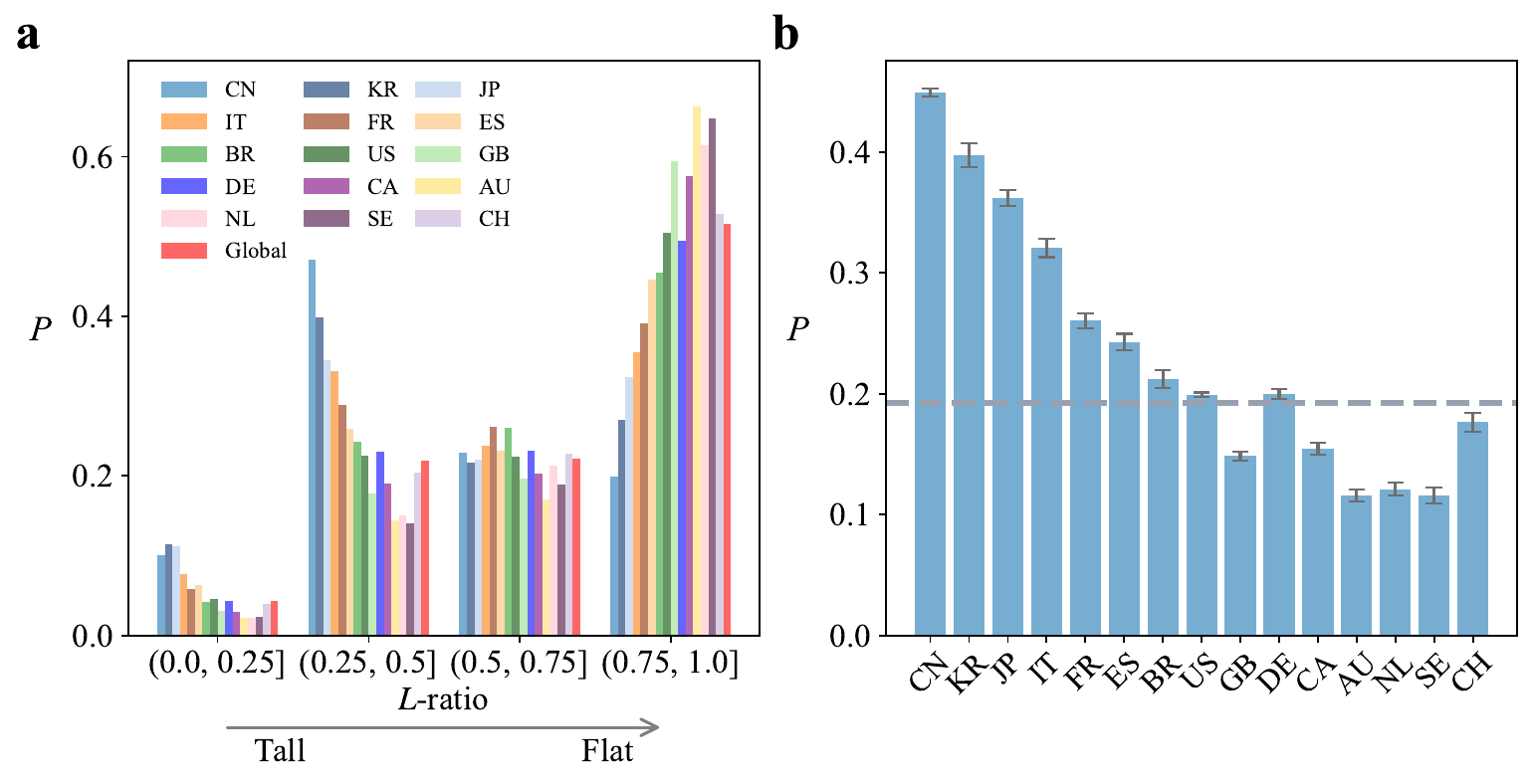}
  \caption{Panel a: Distribution of papers across $L$-ratio intervals for each country, with the red bar representing the global average. Panel b: Proportion of papers produced by tall teams in different countries, with the dashed line representing the global average. Global averages in both panels are calculated after excluding China, Japan, and South Korea. }
\label{fig:fig1}
\end{figure}

We begin by comparing the distribution of $L$-ratio values across the 15 countries with the highest number of publications (Figure \ref{fig:fig1}a). We divide the $L$-ratio into four intervals: (0, 0.25], (0.25, 0.5], (0.5, 0.75], and (0.75, 1.0]. In most countries, the majority of research papers fall within the highest $L$-ratio range ($L$-ratio $>$ 0.75). For example, 50.4\% of papers from the United States and 50.1\% of papers globally fall in this range, reflecting a dominant pattern of flat team structures with distributed leadership. In contrast, flat team structures are considerably less common in China, Japan, and South Korea. In these countries, only 20.0\%, 32.3\%, and 27.0\% of papers, respectively, fall within the highest $L$-ratio range. Instead, they show a marked concentration of papers in the lower $L$-ratio ranges, particularly in the (0.25, 0.5] interval, reflecting their relatively more hierarchical team structures.

To better understand the hierarchical team configurations in each country, we analyze the proportion of tall teams (Figure \ref{fig:fig1}b). China exhibits the highest share of tall teams (45.0\%), followed by South Korea (39.7\%) and Japan (36.2\%). These levels are markedly above both the global average excluding these three countries (20.6\%) and that of the United States (19.9\%). In Europe, Italy (32.0\%), France (26.0\%), and Spain (24.3\%) exhibit high shares of tall teams, but these values remain well below those in China, Japan, and South Korea. Overall, these results highlight the pronounced dominance of tall teams in East Asia.

\begin{figure}[h]
  \centering
  \includegraphics[width= 0.86\textwidth]{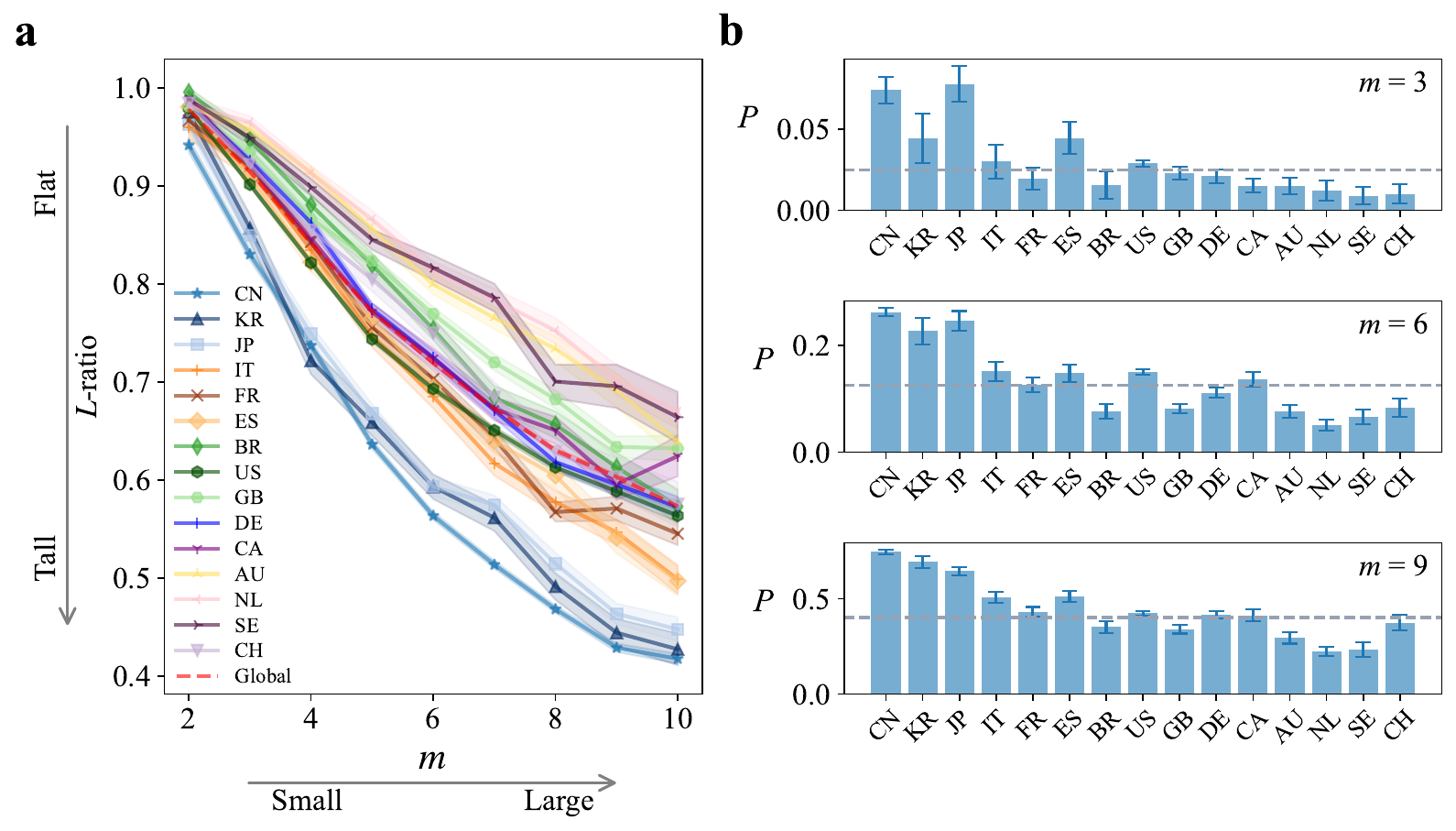}
  \caption{Panel a: Relationship between team size ($m$) and team hierarchy ($L$-ratio) across countries. Panel b: Proportion of papers produced by tall teams across countries for selected team sizes $m =$ 3, 6, 9. In both panels, dashed lines indicate the corresponding global averages, calculated after excluding China, Japan, and South Korea.}
  \label{fig:fig2}
\end{figure}

It is reported that the team hierarchy and size are correlated, as a large team is also likely to be a tall team  \parencite{xu2022flat}. Given that large teams are more frequently observed in scientific outputs from East Asia than other countries worldwide \parencite{liu2021dominance}, it is reasonable to question if the observed dominance of tall teams in East Asia is simply the consequence of the dominance of large teams. To address this question, we first examine the relationship between $L$-ratio and team size for different countries (Figure \ref{fig:fig2}a). As the team size increases, the $L$-ratio decreases in all countries, in line with the previous finding. However, the curves for China, Japan and South Korea lie consistently below those of other countries and the global average, demonstrating a higher degree of hierarchy across nearly all team sizes. Furthermore, the curves for China, Japan and South Korea demonstrate a steeper drop, suggesting that team hierarchy intensifies more significantly with the increasing team size. In other words, a large team from China, Japan and South Korea is more likely to be a tall team than other countries. This effect is most apparent when analyzing the fraction of tall teams when controlling for the team size (Figure \ref{fig:fig2}b and Supplementary Figure S2). In summary, the dominance of big teams in China, Japan and South Korea intensifies the dominance of tall teams in these countries, but the dominance of tall teams cannot be fully explained by the dominance of big teams. There must be some other factors underlying this pattern, which we will discuss later.

\begin{figure}[h]
  \centering
  \includegraphics[width = \textwidth]{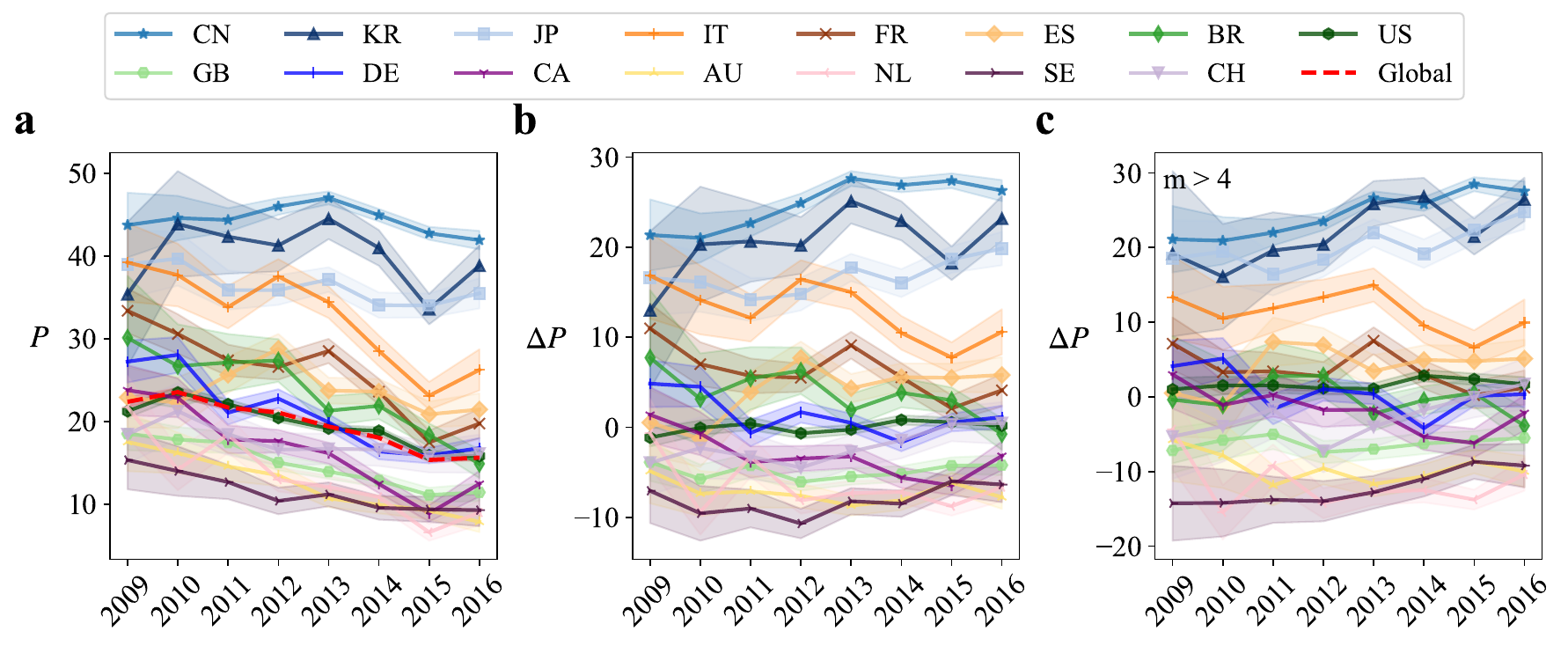}
\caption{Panel a: tall-team proportion (percentage) over time. Panels b and c: deviation from the global average over time, calculated as the annual tall-team proportion in a given country minus the corresponding global average (excluding China, Japan, and South Korea). Panel b shows results without controlling for team size, while panel c restricts to teams with more than four members ($m > 4$).}
\label{fig:fig3}
\end{figure}

Next, we check whether this pattern persists over time. We analyze longitudinal trends in the share of tall teams for each country from 2009 to 2016 and identify a global shift toward flatter teams across all countries (Figure \ref{fig:fig3}a). However, using the global average as the baseline, we observe different extents of change (Figure~\ref{fig:fig3}b). China, Japan, and South Korea are consistently above the global average, demonstrating positive deviations ($\Delta P$) from the baseline. $\Delta P$ of  China exhibits a statistically significant positive slope (0.972, $p = 0.0045$), while Japan (0.520, $p = 0.07$) and South Korea (0.865, $p = 0.14$) have positive but not statistically significant slopes, reflecting a numerical upward trend after taking the global change into consideration. In contrast, most European countries remain close to the global average, with a relatively small value of $\Delta P$. In addition, the slope of $\Delta P$ is close to zero or negative. This observation suggests that while scientific teams in China,  Japan and South Korea are becoming flatter, the change is slower than in other countries worldwide.

Controlling for team size ($m > 4$) reinforces these patterns (Figure~\ref{fig:fig3}c): $\Delta P$ of China (1.161, $p = 0.00036$), Japan (1.249, $p = 0.024$), and South Korea (0.832, $p = 0.029$) all exhibit larger slopes than in the non-controlled case, and all slopes are statistically significant. The pattern persists for other team size thresholds ($m > 6$ and $m > 8$, Supplementary Figure S3). The evolution toward flatter teams is much slower among large teams in China, Japan, and South Korea. Therefore, the trend difference in the three countries is not fully attributed to the dominance of large teams, although the abundance of large teams in these countries can intensify this difference.

We also compute $\Delta P$ using an alternative global baseline. For each focal country, we calculate its annual tall-team proportion and then construct the global baseline by taking the simple average of the tall-team proportions of the other top 14 most productive countries in the same year. Using this alternative baseline, China, Japan, and South Korea continue to exhibit consistently higher tall-team proportions relative to the global level, confirming that our main conclusions are not affected by the choice of global average (Supplementary Figure S4).

\subsection{Possible factors associated with tall-team dominance}

Our analyses reveal a constant pattern that the scientific output from East Asia is dominated by tall teams. This pattern is stable over time and cannot be fully accounted for by team size, indicating that other factors may contribute. To explore potential explanations, we examine two complementary perspectives: (1) Social culture, and (2) scientific funding pattern.

\subsubsection{Social culture: Power Distance}

Social culture constitutes a set of shared values, norms, beliefs, and attitudes that shape behaviors and interactions among members of a society \parencite{schein2010organizational,paais2020effect,tadesse2024organizational,hofstede2011dimensionalizing}. Power Distance (PD), one of its key dimensions, refers to the extent to which less powerful members of a society accept and expect that power is distributed unequally \parencite{hofstede2010cultures,hofstede2011dimensionalizing,hofstede2001culture}. The Power Distance Index (PDI) was originally constructed through ecological factor analysis of country-level averages from large-scale surveys of IBM employees across more than 50 countries \parencite{hofstede1984culture}. Higher PDI values indicate greater acceptance of hierarchical authority, whereas lower values reflect preferences for egalitarianism and decentralized decision-making. Despite concerns regarding cross-country measurement equivalence and response styles \parencite{daniels2014exploring}, PDI remains a stable and widely used indicator for cross-national comparisons of societal hierarchy. Its documented associations with related political and economic structures further support its use at the country level \parencite{gregg1965dimensions,adelman1967society}. Therefore, we use PDI as a cultural lens to explain the dominance of tall teams in East Asia.

\begin{figure}[h]
  \centering
  \includegraphics[width=0.86\textwidth]{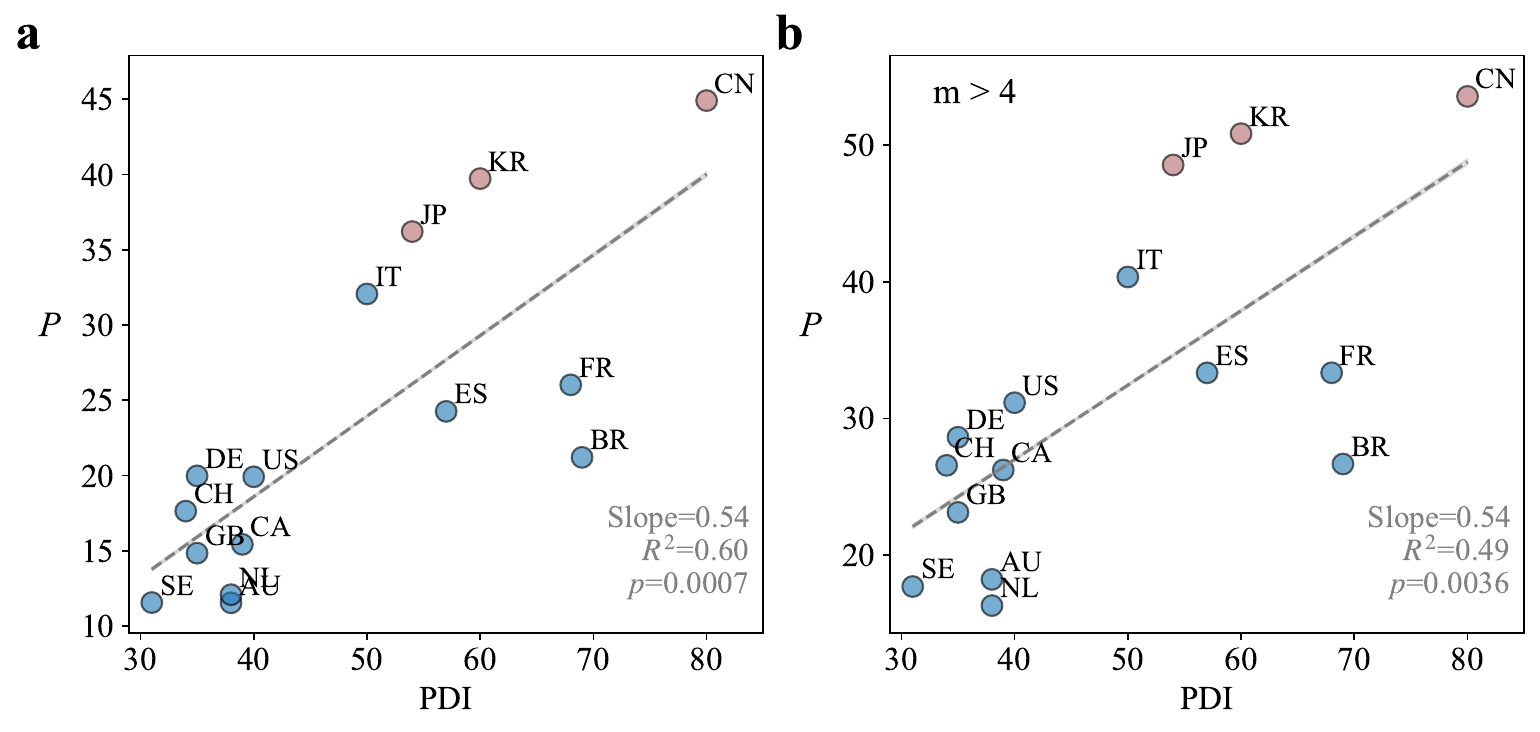}
 \caption{Relationship between national PDI and the proportion of tall teams (expressed as percentage). Panel a shows results without controlling for team size, while panel b restricts to teams with more than four members ($m > 4$). Dashed lines indicate the linear regression fits.}
 \label{fig:fig4}
\end{figure}

We examine the correlation between the proportion of publications produced by tall teams in each country and the corresponding Power Distance Index (PDI) scores (Figure~\ref{fig:fig4}a). Linear regression indicates a strong positive relationship ($\text{slope} = 0.54$, $R^2 = 0.60$, $p = 0.0007$), implying that PDI can explian a substantial portion of cross-national variation in the proportion of tall-team publications. For example, China, with the highest PDI, has the highest proportion of publications produced by tall teams, whereas the United States, the United Kingdom, and Canada have lower PDI and correspondingly have more publications by flat teams. The positive correlation remains after controlling for team size (Figure~\ref{fig:fig4}b and Supplementary Figure S5), with a slope similar to the uncontrolled case and moderately reduced $R^2$, and remains statistically significant. The results provide a reasonable explanation for the dominance of tall teams in East Asia. If the society tends to accept hierarchical organizations, as illustrated by the PDI, the scientific team would not be an exception and its evolution towards more hierachical should be expected.

We also analyze the remaining five Hofstede cultural dimensions (Supplementary Figure S6). Although three of these dimensions show statistically significant associations with the proportion of publications produced by tall teams, PDI is conceptually the most relevant metric for hierarchical team structure and is therefore used as the primary cultural lens to interpret the observed patterns.

\subsubsection{Scientific funding patterns}

A recent study suggests that national funding patterns are associated with typical research team sizes \parencite{liu2021dominance}. Motivated by this observation, we investigate whether they are also associated with the hierarchical structure of research teams. We focus on national funding agencies that primarily support basic or fundamental science in China, Japan, South Korea, and the United States, namely NSFC \parencite{huang2016does}, JSPS \parencite{ohniwa2023effectiveness}, NRF \parencite{jung2017factors}, and NSF \parencite{wang2012science}. This choice is motivated by their broad disciplinary coverage and their pivotal role in shaping national scientific agendas. Moreover, our dataset primarily comprises publications in multidisciplinary journals such as PLOS ONE and PNAS, thereby providing a representative sample of outputs supported by these agencies.

\begin{figure}[h]
	\centering
	\includegraphics[width=\textwidth]{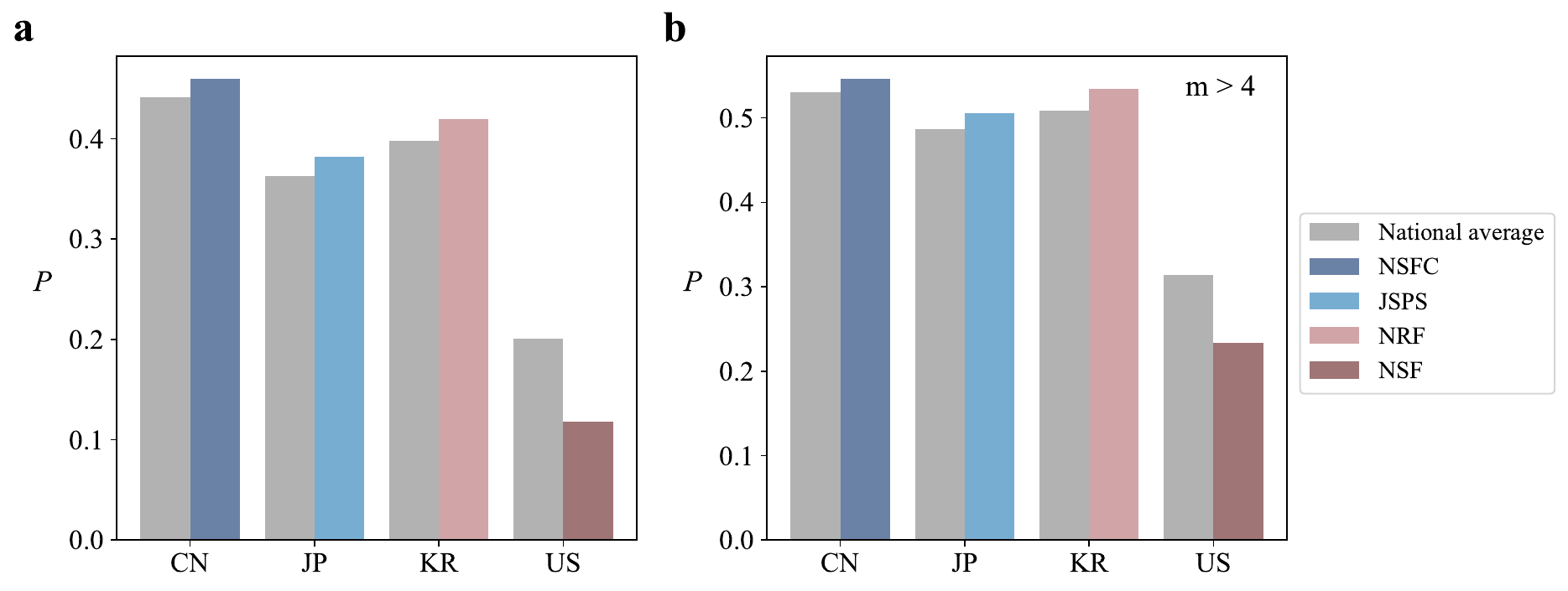}
	\caption{Comparison of the proportion of tall teams in papers supported by the main basic science funding agency in each country (NSFC in China, JSPS in Japan, NRF in South Korea, NSF in the United States) and the national average. Panel a shows results without controlling for team size, while panel b restricts to teams with more than four members ($m > 4$).}
	\label{fig:fig5}
\end{figure}

For each country, we calculate the share of publications produced by tall teams within papers supported by its primary national basic science funding agency, and compare it with the national average share of tall-team publications. In China, Japan, and South Korea, the share of tall teams among papers funded by NSFC, JSPS, and NRF is higher than the national average. In contrast, in the United States, papers funded by NSF exhibit a substantially lower tall-team share than the national average (Figure~\ref{fig:fig5}a). These patterns remain consistent after restricting analyses to papers with $m > 4, 6, 8$ (Figure~\ref{fig:fig5}b and Supplementary Figure S7), indicating that national basic science funding patterns are associated with the proportion of tall-team publications in East Asia.

\subsection{OLS estimates of team hierarchy correlates: region, culture, and funding}

Building on the descriptive evidence that East Asian teams exhibit more pronounced hierarchy (lower $L$-ratios) and a steeper increase of hierarchy with team size, we now quantify the associated factors through paper-level OLS regressions. We examine three distinct dimensions separately: regional affiliation, societal hierarchy culture, and national funding patterns. In all models, the dependent variable is the paper-level $L$-ratio. The key explanatory variables are, respectively, an East Asia indicator (EA), the national PDI, and the share of tall-team papers funded by the primary national agency ($f$). Team size ($m$) is included as a core control, along with its interaction with each main explanatory variable to test whether the relationship between size and hierarchy is moderated by region, culture, or funding.

\begin{table}[htbp]
\centering
\caption{Regression analysis of factors associated with East Asian team hierarchy. 
Nine OLS models estimate the effects of a regional indicator (EA), a cultural dimension (PDI), and national funding patterns ($f$) on the paper-level $L$-ratio. Each dimension is examined with progressive controls for $m$ and interaction terms. Each regression coefficient is tested against the null hypothesis that the coefficient equals zero using a two-sided t-test. Robust standard errors clustered at the country level are shown in parentheses. Significance: *** \(p<0.001\), ** \(p<0.01\), * \(p<0.05\).}
\label{tab:three_dimensions}
\begin{threeparttable}
\tiny
\begin{tabular}{lccccccccc}
\toprule
& \multicolumn{3}{c}{\textbf{Region}} & \multicolumn{3}{c}{\textbf{Culture}} & \multicolumn{3}{c}{\textbf{Funding}} \\
\cmidrule(lr){2-4} \cmidrule(lr){5-7} \cmidrule(lr){8-10}
& \textbf{Model 1} & \textbf{Model 2} & \textbf{Model 3} & \textbf{Model 4} & \textbf{Model 5} & \textbf{Model 6} & \textbf{Model 7} & \textbf{Model 8} & \textbf{Model 9} \\
\midrule
~~EA & $-0.197$*** & $-0.166$*** & $-0.162$*** & & & & & & \\
~~ & (0.019) & (0.016) & (0.016) & & & & & & \\
~~EA × m & & & $-0.034$*** & & & & & & \\
~~ & & & (0.006) & & & & & & \\
~~PDI & & & & $-0.080$*** & $-0.065$*** & $-0.064$*** & & & \\
~~ & & & & (0.008) & (0.008) & (0.008) & & & \\
~~PDI × m & & & & & & $-0.012$*** & & & \\
~~ & & & & & & (0.003) & & & \\
~~f
  & & & & & & & $-0.082$*** & $-0.068$*** & $-0.066$*** \\
~~ & & & & & & & (0.011) & (0.009) & (0.009) \\
~~f × m & & & & & & & & & $-0.015$*** \\
~~ & & & & & & & & & (0.003) \\
~m & & $-0.129$*** & $-0.123$*** & & $-0.129$*** & $-0.130$*** & & $-0.129$*** & $-0.131$*** \\
~~ & & (0.007) & (0.005) & & (0.007) & (0.004) & & (0.007) & (0.004) \\
\midrule
~~Observations & {121,197} & {121,197} & {121,197} & {121,197} & {121,197} & {121,197} & {121,197} & {121,197} & {121,197} \\
~~Adjusted \(R^2\) & 0.165 & 0.355 & 0.357 & 0.160 & 0.348 & 0.350 & 0.164 & 0.352 & 0.354 \\
\bottomrule
\end{tabular}
\end{threeparttable}
\end{table}

Table~\ref{tab:three_dimensions} presents the regression results. Three key patterns emerge. First, regarding regional affiliation, EA exhibits a significant negative association with the $L$-ratio ($-$0.197 in Model 1). This corresponds to a substantial and statistically significant hierarchy gap between East Asian and non-East Asian teams. Controlling for $m$ in Model 2 reduces the magnitude of this effect only modestly (to $-$0.166), indicating that differences in average $m$ explain a small portion of the regional gap. Furthermore, the significant negative interaction between EA and $m$ in Model 3 ($-$0.034) reveals that hierarchy intensifies more sharply as $m$ increases within East Asia compared to other regions. Second, for cultural context, a one-standard-deviation increase in PDI is associated with a 0.080 decrease in the $L$-ratio (Model 4), signaling more hierarchical teams. This relationship remains robust after controlling for $m$ (Model 5, -0.065). The negative interaction with $m$ in Model 6 suggests that the hierarchical effect of larger teams is amplified in high-PDI cultural settings. Third, examining funding patterns, a one-standard-deviation increase in $f$ is linked to a 0.082 decrease in the $L$-ratio (Model 7). This association is also largely independent of $m$ (Model 8, -0.068) and features a significant negative interaction with $m$ in Model 9, paralleling the findings for culture.

\begin{table}[htbp]
\centering
\caption{Predicted $\Delta L$-ratios between East Asian and non-East Asian teams, calculated based on differences in $m$, PDI, and $f$.}
\label{tab:ea_prediction}
\begin{threeparttable}
\scriptsize
\begin{tabular}{lccc}
\toprule
 & \textbf{PDI} & \textbf{f} & \textbf{$m$} \\
\midrule
East Asia (z-score) 
& 1.521 & 1.721 & 0.173 \\
Non-East Asia (z-score) 
& $-0.444$ & $-0.502$ & $-0.051$ \\
Difference (East Asia -- Non-East Asia) 
& 1.965 & 2.223 & 0.224 \\
\midrule
Standardized regression coefficient 
& $-0.0653$ (Model 5) & $-0.0678$ (Model 8) & $-0.1293$ (Model 2)\\
\midrule
Predicted $\Delta L$-ratio 
& $-0.0653$*1.965 = $-0.128$ & $-0.0678$*2.223 = $-0.151$ & $-0.1293$*0.224 = $-0.029$ \\
\bottomrule
\end{tabular}
\end{threeparttable}
\end{table}

To assess how much of the observed East Asian hierarchy gap can be independently associated with cultural and funding patterns, we estimate the potential contribution of each factor separately. The coefficient of EA from Model 2 (-0.166, Table~\ref{tab:three_dimensions}) serves as a benchmark for the total observed gap after controlling for $m$. For each factor, we calculate a predicted $\Delta L$-ratio by multiplying its standardized regression coefficient (from Models 5 and 8 for PDI and f, respectively) by the standardized mean difference in that variable between East Asian and non–East Asian teams. As shown in Table~\ref{tab:ea_prediction}, PDI predicts a $\Delta L$-ratio of -0.128 and f predicts -0.151, while differences in average $m$ account for only -0.029. The predicted contribution of either PDI or f is close to the total observed gap of -0.166. This indicates that either cultural context or national funding patterns, on its own, can explain a large majority of the hierarchy gap between East Asian and non-East Asian teams.

\section{Conclusion}

This study investigates the hierarchical configuration of scientific teams in different countries, drawing on 150,817 publications with author contribution statements. Among the 15 countries with the highest number of publications, China, Japan, and South Korea consistently exhibit a higher proportion of tall teams. Although there is a strong correlation between team size and hierarchy, we find that the dominance of tall teams in China, Japan, and South Korea cannot be fully explained by the prevalence of large teams in these countries. A large team in these countries is also more likely to be a tall team, compared with other countries worldwide. In other words, the dominance of large teams in East Asia intensifies the dominance of tall teams. But the dominance of tall teams would emerge on its own, independent of team size. In addition, when considering temporal trends, we find that scientific teams are becoming flatter in all countries. However, this trend in China, Japan, and South Korea is slower compared with the global average. This slower pace contributes to the higher proportion of tall teams in East Asia, as tall teams persist longer in the region. To contextualize this persistent pattern, we examine two factors: social culture and funding patterns. Countries with higher Power Distance Index (PDI) tend to have more tall teams, which contributes to the pronounced dominance of tall teams in China, Japan, and South Korea. In addition, the primary basic science funding agencies in these countries, the National Natural Science Foundation of China (NSFC), the Japan Society for the Promotion of Science (JSPS), and the National Research Foundation of Korea (NRF), support taller teams relative to their national averages, whereas the United States National Science Foundation (NSF) shows the opposite trend. Together, these cultural and funding factors help account for the sustained dominance of tall teams in East Asia. In addition, paper-level regression analyses controlling for team size, citation impact, international collaboration, year, and field fixed effects show that East Asian teams continue to exhibit stronger hierarchy compared with non-East Asian teams, with PDI and the share of tall-team papers funded by major basic science agencies predicting this hierarchy gap more strongly than team size.

Our study makes the following contributions. First, it provides a systematic cross-national analysis of team hierarchy, revealing a distinctive East Asian pattern in both static and temporal dimensions. Second, we conduct a mechanism-oriented analysis, linking this distinctive hierarchy to national cultural factors and funding patterns, offering a mechanistic understanding of cross-national differences in team structure. Beyond these academic contributions, our findings also offer insights for the management and design of scientific teams. A recent study has highlighted that the scarcity of small teams may constrain the generation of highly disruptive innovations \parencite{liu2021dominance}. Our finding that tall teams dominate scientific output in East Asia may similarly limit disruptive innovation, given that flat teams have been shown to exhibit higher innovation capacity \parencite{xu2022flat}. Since this predominance of tall teams persists independently of team size, adjusting team hierarchy by promoting shared leadership and more distributed responsibilities may be a more effective strategy for enhancing innovation potential than focusing solely on team size. At the same time, the growing complexity of scientific challenges necessitates large collaborative teams, making careful hierarchical design essential for balancing team efficiency and creative capacity.

Interestingly, prior research suggests that different definitions of leadership yield opposite conclusions about the relationship between team hierarchy and innovation. When hierarchy is assessed through shared leadership, which encompasses not only participation in research conception but also responsibilities such as management, supervision, and coordination, flatter teams tend to generate more disruptive ideas \parencite{xu2022flat}. In contrast, when measured by thought leadership, which focuses primarily on involvement in research conception and design, flatter teams may actually suppress disruption \parencite{zhao2024more}. Our analysis shows that under both leadership definitions, flat teams are consistently and significantly underrepresented in East Asian countries relative to the global average. These patterns highlight a critical consideration: while increasing the number of lead authors can help flatten team structures at a given team size, whether this should be achieved by expanding thought leadership or shared managerial leadership requires careful deliberation, as each may have different implications for innovation.

Prior work suggests that flatter teams, characterized by more equally shared intellectual and managerial responsibilities, tend to foster greater scientific disruption \parencite{xu2022flat}, possibly because shared leadership may facilitate more flexible recombination of knowledge within the team \parencite{yu2025knowledge}. However, the cross-national generalizability of the relationship between team hierarchy and disruption remains unclear. Country-specific research is an interesting and widely studied direction, highlighting how national contexts shape scientific practices and outputs \parencite{li2025affiliation, shi2023has, wang2022novel, alam2025perceptions, huang2024talent}. Future work could systematically investigate how hierarchical structures interact with national and institutional factors to guide effective team design and policy interventions.

This study has several limitations that outline important directions for future work. First, although field fixed effects account for average differences across disciplines, the estimated association primarily reflects a sample-size–weighted average dominated by Biology and Medicine, which constitute the majority of the sample. Whether the same pattern holds within less-represented disciplines remains unclear and warrants further investigation. Second, while we focus on cultural and funding factors, other contextual influences such as institutional policies may also affect team hierarchy but are not examined here. Third, the $L$-ratio is a measure of leadership concentration, corresponding to the centralization dimension within the inequality perspective of hierarchy. It does not address other hierarchical dimensions like steepness or acyclicity.

\section*{Funding information}

This work was supported by grants from the Natural Science Foundation of China (No. 72374173), the University Innovation Research Group of Chongqing (No. CXQT21005), and the Fundamental Research Funds for the Central Universities (No. SWU-XDJH202303).

\section*{Author contributions}

Siyuan Liu (liusiyuan311@163.com; ORCID: 0000-0002-4106-7135): Conceptualization; Methodology; Data curation; Formal analysis; Writing - Original Draft. 

Wenjin Xie (xiewenjin@email.swu.edu.cn; ORCID: 0000-0002-1438-8443): Formal analysis. 

Wenyu Chen (shanyu618@163.com; ORCID: 0000-0003-4299-8764): Formal analysis. 

Tao Jia (tjia@swu.edu.cn; ORCID: 0000-0002-2337-2857): Conceptualization; Formal analysis; Funding acquisition; Writing – review \& editing.

\section*{Data availability statements}

The data that support the findings of this study are openly
available in Science Data Bank at https://www.doi.org/10.57760/sciencedb.33995.

\printbibliography

\end{document}